\documentclass[runningheads]{llncs}

\usepackage{times}
\usepackage{helvet}
\usepackage{courier}
\usepackage{setspace}

\usepackage{color}
\usepackage{comment}
\usepackage{url}

\usepackage{booktabs}
\usepackage{amsfonts}
\usepackage{amsmath}
\usepackage{amssymb}
\usepackage{makecell}
\usepackage{multirow}
\usepackage{pbox}

\usepackage[flushleft]{threeparttable}

\usepackage{graphicx,xspace}
\usepackage{epstopdf}
\usepackage{stfloats}
\usepackage[font=small]{caption}
\usepackage{enumitem}
\usepackage{wrapfig}
\def\st#1{~}
\newcommand{\dsnamens}{WAP196k}
\newcommand{\dsname}{{\dsnamens}\ }
\newcommand{\ggnamens}{Google}

\begin{document}
\title{Neural Article Pair Modeling for Wikipedia\\ Sub-article Matching\thanks{This work is accomplished at Google, Mountain View.}}

%\author{
%Muhao Chen$^1$, Changping Meng$^2$, Gang Huang$^3$,Carlo Zaniolo$^1$\\
%Department of Computer Science, University of California, Los Angeles$^1$\\
%Department of Computer Science, Purdue University, West Lafayette$^3$\\
%Google Inc., Mountain View$^3$\\
%\{muhaochen, zaniolo\}@cs.ucla.edu, meng40@purdue.edu, ganghuang@google.com
%}
\author{Muhao Chen\inst{1} \and
Changping Meng\inst{2} \and
Gang Huang\inst{3} \and
Carlo Zaniolo\inst{1}}
\institute{University of California, Los Angeles, CA, USA\\
\email{\{muhaochen, zaniolo\}@cs.ucla.edu}\\
\and
Purdue University, West Lafayette, IN, USA\\
\email{meng40@purdue.edu}\\
\and
Google, Mountain View, CA, USA\\
\email{ganghuang@google.com}
}
\maketitle

\setcounter{footnote}{0}
%\begin{abstract}
%Nowadays, editors tend to separate different subtopics of a long Wikipedia article into multiple {\em sub-articles}.
%Such split-off of contents, although improves the readability, is however problematic for many Wikipedia-based techniques and corresponding applications that rigidly follow the \emph{article-as-concept} assumption, which requires each entity (or concept) to be described solely by one article.
%These techniques include but are not limited to knowledge base construction, cross-lingual knowledge alignment, semantic search and data lineage of Wikipedia entities.
%Thus, {\em sub-article matching} is highly desired for Wikipedia, which helps restore the scattered pieces of knowledge to whole views, and better support the above Wikipedia-based techniques.
%In this paper, we propose a neural article pair model to address the sub-article matching problem.
%The model adopts a hierarchical learning structure that combines variants of neural document encoders with a comprehensive set of explicit features.
%To generalize the problem, a large dataset is created via massive crowdsourcing and strategic rules.
%The proposed model achieves promising results of cross-validation and significantly outperforms previous approaches on the large dataset.
%Serving of the proposed model on the entire English Wikipedia also indicates it to be beneficial to the construction of a large-scale knowledge base by effectively extracting a vast collection of main and sub-article pairs.
%\end{abstract}
\begin{abstract}
Nowadays, editors tend to separate different subtopics of a long Wiki-pedia article into multiple {\em sub-articles}.
This separation seeks to improve human readability.
However, it also has a deleterious effect on many Wikipedia-based tasks that rely on the \emph{article-as-concept} assumption, which requires each entity (or concept) to be described solely by one article.
This underlying assumption significantly simplifies knowledge representation and extraction, and it is vital to many existing technologies such as automated knowledge base construction, cross-lingual knowledge alignment, semantic search and data lineage of Wikipedia entities.
In this paper we provide an approach to match the scattered sub-articles back to their corresponding main-articles, with the intent of facilitating automated Wikipedia curation and processing.
The proposed model adopts a hierarchical learning structure that combines multiple variants of neural document pair encoders with a comprehensive set of explicit features.
A large crowdsourced dataset is created to support the evaluation and feature extraction for the task.
Based on the large dataset, the proposed model achieves promising results of cross-validation and significantly outperforms previous approaches.
Large-scale serving on the entire English Wikipedia also proves the practicability and scalability of the proposed model by effectively extracting a vast collection of newly paired main and sub-articles.
\end{abstract}

\section{Introduction}
Wikipedia has been the essential source of knowledge
for people as well as computing research and practice.
This vast storage of encyclopedia articles for real-world entities (or concepts) has brought along the automatic construction of knowledge bases \cite{lehmann2015dbpedia,mahdisoltani2014yago3} that support knowledge-driven computer applications with vast structured knowledge.
Meanwhile, Wikipedia also triggers the emerging of countless AI-related technologies for semantic web  \cite{vrandevcic2014wikidata,meij2014entity,zou2014natural}, natural language understanding \cite{ni2016semantic,chen2017reading,wang2012cross}, content retrieval \cite{ackerman2013sharing,kittur2010beyond}, and other aspects.\par

\begin{figure}[t]
  \centering
  \includegraphics[width=1.0\columnwidth]{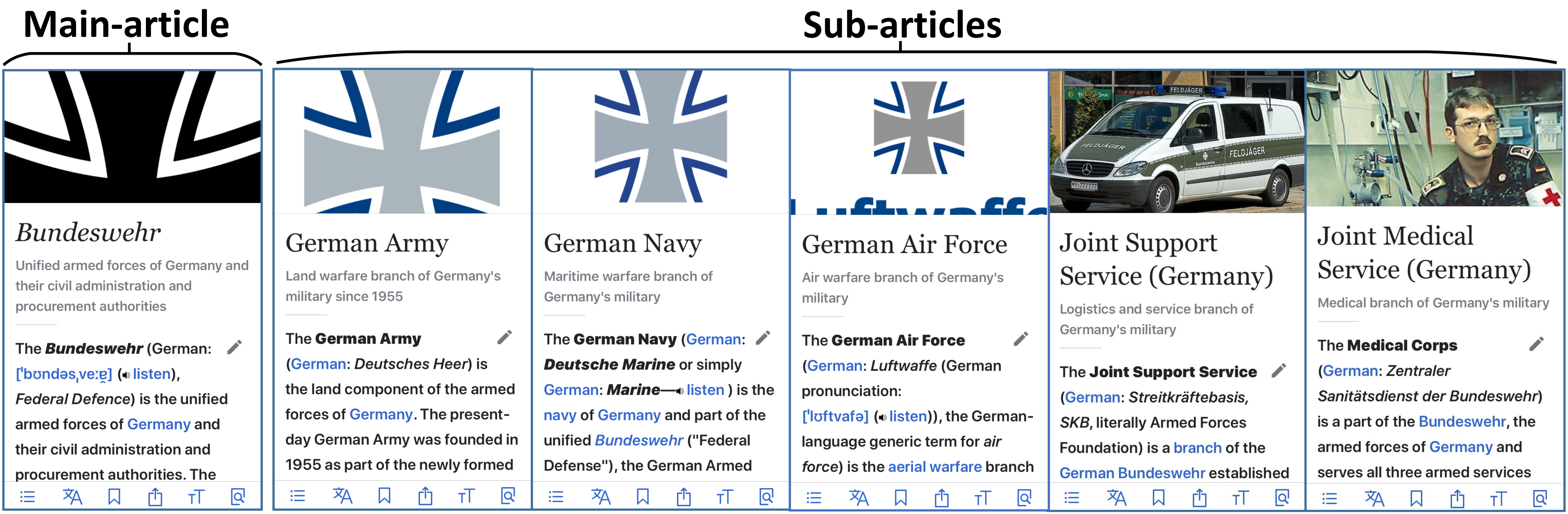}\\
  \caption{The main-article \emph{Bundeswehr} (unified armed forces of Germany) and its sub-articles \emph{German Army}, \emph{German Navy}, \emph{German Airforce}, \emph{Joint Support Service of Germany} and \emph{Joint Medical Service of Germany}.}
  \label{fig:bundeswehr}
  \vspace{-1.5em}
\end{figure}

Most existing automated Wikipedia techniques assume the one-to-one mapping between entities and Wikipedia articles \cite{lin2017problematizing,dojchinovski2013entityclassifier,mahdisoltani2014yago3}.
This so-called {\em article-as-concept} assumption \cite{lin2017problematizing} regulates each entity to be described by at most one article in a language-specific version of Wikipedia. %as encourage by the recent Wikipedia guidelines \cite{url:wikisize},
However, recent development of Wikipedia itself is now breaking this assumption, as rich contents of an entity are likely to be separated in different articles and managed independently.
For example, many details about the entity ``Harry Potter'' are contained in other articles such as ``Harry Potter Universe'', ``Harry Potter influences and analogues'', and ``Harry Potter in translation''.
Such separation of entity contents categorizes Wikipedia articles into two groups:
the {\em main-article} that summarizes an entity, and the {\em sub-article} that comprehensively describes an aspect or a subtopic of the main-article.
Consider another example: for the main-article \emph{Bundeswehr} (i.e. unified armed forces of Germany) in English Wikipedia, we can find its split-off sub-articles such as \emph{German Army}, \emph{German Navy}, \emph{German Airforce}, \emph{Joint Support Service of Germany}, and \emph{Joint Medical Service of Germany} (as shown in Fig.~\ref{fig:bundeswehr}).
This type of sub-article splitting is quite common on Wikipedia.
Around 71\% of the most-viewed Wikipedia entities are split-off to an average of 7.5 sub-articles \cite{lin2017problematizing}.
\par

While sub-articles may enhance human readabilities, the violation of the article-as-concept assumption is however problematic to a wide range of Wikipedia-based technologies and applications that critically rely on this assumption.
%In fact the recognition among main-articles and corresponding sub-articles are usually far from complete, as many of these articles are created and evolve independently \cite{lin2017problematizing}.
For instance, Wikipedia-based knowledge base construction \cite{lehmann2015dbpedia,mahdisoltani2014yago3,mousavi2014text} will assume the article title as an entity name, which is then associated with the majority of relation facts for the entity from the infobox of the corresponding article.
Split-off of articles sow confusion in a knowledge base extracted with these existing techniques.
Clearly, explicit \cite{gabrilovich2007computing,hecht2012explanatory} and implicit semantic representation techniques \cite{schuhmacher2014knowledge,xie2016representation,chen2017learning} based on Wikipedia are impaired due to that a part of links and text features utilized by these approaches are now likely to be isolated from the entities, which further affects NLP tasks based on these semantic representation techniques such as semantic relatedness analysis \cite{strube2006wikirelate,liu2016cross,ni2016semantic}, relation extraction \cite{mousavi2014text,chen2018onto}, and named entity disambiguation \cite{yamada2016joint}.
Multilingual tasks such as knowledge alignment \cite{chen2017multigraph,chen2018co,suchanek2011paris,wang2012cross} and cross-lingual Wikification \cite{tsai2016cross} become challenging for entities with multiple articles, since these tasks assume that we have a one-to-one match between articles in both languages.
Semantic search of entities \cite{meij2014entity,zou2014natural,cai2013wikification} is also affected due to the diffused nature of the associations between entities and articles.\par

To support the above techniques and applications, it is vital to address the \emph{sub-article matching} problem, which aims to restore the complete view of each entity by matching the sub-articles back to the main-article.
However, it is non-trivial to develop a model which recognizes the implicit relations that exist between main and sub-articles.
A recent work \cite{lin2017problematizing} has attempted to tackle this problem by characterizing the match of main and sub articles using some explicit features.
These features focus on measuring the symbolic similarities of article and section titles, the structural similarity of entity templates and page links, as well as cross-lingual co-occurrence.
Although these features are helpful to identify the sub-article relations among a small scale of article pairs, they are still far from fully characterizing the sub-article relations in the large body of Wikipedia.
And more importantly, the semantic information contained in the titles and text contents, which is critical to the characterization of semantic relations of articles, has not been used for this task.
However, effective utilization of the semantic information would also require a large collection of labeled main and sub-article pairs to generalize the modeling of such implicit features.\par

In this paper, we introduce a new approach for addressing the sub-article matching problem.
Our approach adopts neural document encoders to capture the semantic features from the titles and text contents of a candidate article pair, for which several encoding techniques are explored with.
Besides the semantic features, the model also utilizes a set of explicit features that measure the symbolic and structural aspects of an article pair.
Using a combination of these features, the model decides whether an article is the sub-article of another.
To generalize the problem, massive crowdsourcing and strategic rules are applied to create a large dataset that contains around 196k Wikipedia article pairs, where around 10\% are positive matches of main and sub-articles, and the rest comprehensively cover different patterns of negative matches.
Held-out estimation proves effectiveness of our approach by significantly outperforming previous baselines, and reaching near-perfect precision and recall for detecting positive main and sub-article matches from all candidates.
To show the practicability of our approach, we also employ our model to extract main and sub-article matches in the entire English Wikipedia using a 3,000-machine MapReduce \cite{dean2008mapreduce}.
This process has produced a large collection of new high-quality main and sub-article matches, and are now being migrated into a large-scale production knowledge base.\par

The rest of the paper is organized as follows. We first discuss the related work, then introduce our approach in the section that follows. After that, we present the experimental results, and conclude the paper in the last section.

\newcommand{\stitle}[1]{\vspace{0.3ex}\noindent{\bf #1}}

%\inv
\section{Related Work} \label{sec:related}
A recent work \cite{lin2017problematizing} has launched the first attempt to address the Wikipedia sub-article matching problem, in which the authors have defined the problem into the binary classification of candidate article pairs.
Each article pair is characterized based on a group of explicit features that lies in three categories: (1) symbolic similarity: this includes token overlapping of the titles, maximum token overlapping among the section titles of the two articles, and term frequencies of the main-article titles in the candidate sub-article contents; (2) structural similarity: this includes structure similarity of article templates, link-based centrality measures and the Milne-Witten Index \cite{milne2008learning}; (3) cross-lingual co-occurrence: these features consider the proportion of languages where the given article pairs have been identified as main and sub-articles, and the relative multilingual ``globalness'' measure of the candidate main-article.
Although some statistical classification models learnt on these explicit features have offered satisfactory accuracy of binary classification on a small dataset of 3k article pairs that cover a subset of the most-viewed Wikipedia articles, such simple characterization is no-doubt far from generalizing the problem.
When the dataset scales up to the range of the entire Wikipedia, it is very easy to find numerous counterfactual cases for these features.
Moreover, the cross-lingual co-occurrence-based features are not generally usable due to the incompleteness of inter-lingual links that match the cross-lingual counterparts of Wikipedia articles.
Some recent works have even pointed out that such cross-lingual alignment information only covers less than 15\% of the articles \cite{chen2017multigraph,lehmann2015dbpedia,vrandevcic2012wikidata}.
More importantly, we argue that the latent semantic information of the articles should be captured, so as to provide more generalized and comprehensive characterization of the relation for each article pair.
\par

Sentence or article matching tasks such as textual entailment recognition \cite{sha2016reading,yin2016abcnn,hu2014convolutional} and paraphrase identification \cite{yin2015convolutional,yin2016abcnn} require the model to identify content-based discourse relations of sentences or paragraphs \cite{lascarides1993temporal}, which reflect logical orders and semantic consistency.
Many recent efforts adopt different forms of deep neural document encoders to tackle these tasks, where several encoding techniques have been widely employed, including convolutional neural networks \cite{hu2014convolutional,yin2015convolutional}, recurrent neural networks \cite{sha2016reading}, and attentive techniques \cite{yin2016abcnn,rocktaschel2015reasoning,kadlec2016text}.
Detection of the sub-article relations requires the model to capture a high-level understanding of both contents and text structuring of articles.
Unlike the previously mentioned discourse relations, the sub-article relations can be reflected from different components of Wikipedia articles including titles, text contents and link structures.
To tackle new and challenging task of sub-article matching, we incorporate neural document encoders with explicit features in our model, so as to capture the sub-article relation based on both symbolic and semantic aspects of the Wikipedia article pairs.
Meanwhile, we also take efforts to prepare a large collection of article pairs that seek to well generalize the problem.

\def\kb{\mathit{KB}}
\def\bhline{\specialrule{.2em}{0em}{0em}}
\newcommand{\bigO}[1]{{\rm O} (#1)\xspace}

\section{Modeling}

In this section, we introduce the proposed model for the Wikipedia sub-article matching task.
We begin with the denotations and problem definition.\par

\subsection{Preliminaries}\label{sect:prelim}
\stitle{Denotations.} We use $W$ to denote the set of Wikipedia articles, in which we model an article $A_i\in W$ as a triple $A_i=(t_i,c_i,s_i)$.
$t_i$ is the title, $c_i$ the text contents, and $s_i$ the miscellaneous structural information such as templates, sections and links.
$t_i=\{w_{t1}, w_{t2}, ..., w_{tl}\}$ and $c_i=\{w_{c1}, w_{c2}, ..., w_{cm}\}$ thereof are both sequences of words.
In practice, we use the first paragraph of $A_i$ to represent $c_i$ since it is the summary of the article contents.
For each word $w_{i}$, we use bold-faced $\mathbf{w}_{i}$ to denote its embedding representation.
We use $F(A_i,A_j)$ to denote a sequence of explicit features that provide some symbolic and structural measures for titles, text contents and link structures, which we are going to specify in Section \ref{sec:explicit_features}.
We assume that all articles are written in the same language, as it is normally the case of main-articles and their sub-articles.
In this paper, we only consider English articles \emph{w.l.o.g}.
\par
\stitle{Problem definition.}
\emph{Sub-article matching} is defined as a binary classification problem on a set of candidate article pairs $P\subseteq W\times W$.
Given a pair of articles $p=(A_i, A_j)\in P$, a model should decide whether $A_j$ is the sub-article of $A_i$.
The problem definition complies with the previous work that first introduces the problem \cite{lin2017problematizing},
and is related to other sentence matching problems for discourse relations such as text entailment recognition and paraphrase identification \cite{yin2015convolutional,sha2016reading,poria2015deep}.\par
The sub-article relation is qualified based on two criteria, i.e.
$A_j$ is a sub-article of $A_i$ if (1) $A_j$ describes an aspect or a subtopic of $A_i$, and (2) $c_j$ can be inserted as a section of $A_i$ without breaking the topic summarized by $t_i$.
It is noteworthy that the sub-article relation is anti-symmetric, i.e. if $A_j$ is a sub-article of $A_i$ then $A_i$ is not a sub-article of $A_j$.
We follow these two criteria in the crowdsourcing process for dataset creation, as we are going to explain in Section \ref{sect:data}.
To address the sub-article matching problem, our model learns on a combination of two aspects of the Wikipedia articles.
Neural document encoders extract the implicit semantic features from text, while a series of explicit features are incorporated to characterize the symbolic or structural aspects.
In the following, we introduce each component of our model in detail.

\subsection{Document Encoders}

A neural document encoder $E(X)$ encodes a sequence of words $X$ into a latent vector representation of the sequence.
We investigate three widely-used techniques for document encoding \cite{hu2014convolutional,yin2015convolutional,sha2016reading,yin2016abcnn,rocktaschel2015reasoning}, which lead to three types of encoders for both titles and text contents of Wikipedia articles, i.e. convolutional encoders (CNN), gated recurrent unit encoders (GRU), and attentive encoders.

\subsubsection{Convolutional Encoders}
A convolutional encoder employs the 1-dimensional weight-sharing convolution layer to encode an input sequence.
Given the input sequence $X=\{x_1,x_2,...,x_l\}$, a convolution layer applies a kernel $\mathbf{M}_c\in \mathbb{R}^{h \times k}$ to generate a latent representation $\mathbf{h}^{(1)}_i$ from a window of the input vector sequence $\mathbf{x}_{i:i+h-1}$ by

\begin{equation*}
\mathbf{h}^{(1)}_i=\mathrm{tanh}(\mathbf{M}_c \mathbf{x}_{i:i+h-1} + \mathbf{b}_c)
\end{equation*}

for which $h$ is the kernel size and $\mathbf{b}_c$ is a bias vector.
The convolution layer applies the kernel to all consecutive windows to produce a sequence of latent vectors $\mathbf{H}^{(1)}=\{ \mathbf{h}^{(1)}_1, \mathbf{h}^{(1)}_2, ..., \mathbf{h}^{(1)}_{l-h+1}\}$,
where each latent vector leverages the significant local semantic features from each $h$-gram of the input sequence.
Like many other works~\cite{kim2014convolutional,hu2014convolutional,severyn2015twitter}, we apply $n$-max-pooling to extract robust features from each $n$-gram of the convolution outputs by $\mathbf{h}^{(2)}_{i}=\max ( \mathbf{h}^{(1)}_{i:n+i-1})$.\par

\subsubsection{Gated Recurrent Unit Encoder}
The gated recurrent unit (GRU) is an alternative of the long-short-term memory network (LSTM) that has been popular for sentence (sequence) encoders in recent works \cite{jozefowicz2015empirical,dhingra2017gated}.
Each unit consists of two types of gates to track the state of sequences without using separated memory cells, i.e. the reset gate $\mathbf{r}_i$ and the update gate $\mathbf{z}_i$.
%We follow the definition of GRU in~\cite{yang2016hierarchical}.
Given the vector representation $\mathbf{x}_i$ of an incoming item $x_i$, GRU updates the current state $\mathbf{h}^{(3)}_i$ of the sequence as a linear interpolation between the previous state $\mathbf{h}^{(3)}_{i-1}$ and the candidate state $\tilde{\mathbf{h}}^{(3)}_i$ of new item $x_i$, which is calculated as below.

\begin{equation*}
\mathbf{h}^{(3)}_i=\mathbf{z}_i\odot \tilde{\mathbf{h}}^{(3)}_i+(1-\mathbf{z}_i)\odot \mathbf{h}^{(3)}_{i-1}
\end{equation*}

The update gate $\mathbf{z}_i$ that balances between the information of the previous sequence and the new item is updated as below,
where $\mathbf{M}_z$ and $\mathbf{N}_z$ are two weight matrices, $\mathbf{b}_z$ is a bias vector, and $\sigma$ is the sigmoid function.

\begin{equation*}
\mathbf{z}_i=\sigma\left (\mathbf{M}_z\mathbf{x}_i+\mathbf{N}_z \mathbf{h}^{(3)}_{i-1} + \mathbf{b}_z\right )
\end{equation*}

The candidate state $\tilde{\mathbf{h}}^{(3)}_i$ is calculated similarly to those in a traditional recurrent unit as below, where $\mathbf{M}_s$ and $\mathbf{N}_s$ are two weight matrices, and $\mathbf{b}_s$ is a bias vector.

\begin{equation*}
\tilde{\mathbf{h}}^{(3)}_i=\mathrm{tanh}\left (\mathbf{M}_s\mathbf{x}_i+\mathbf{r}_i\odot(\mathbf{N}_s \mathbf{h}^{(3)}_{i-1}) + \mathbf{b}_s\right )
\end{equation*}

The reset gate $\mathbf{r}_i$ thereof controls how much information of the past sequence contributes to the candidate state

\begin{equation*}
\mathbf{r}_i=\sigma\left (\mathbf{M}_r\mathbf{x}_i+\mathbf{N}_r \mathbf{h}^{(3)}_{i-1} + \mathbf{b}_r\right )
\end{equation*}

The above defines a GRU layer
which outputs a sequence of hidden state vectors given the input sequence $X$.
Unlike CNN that focuses on leveraging local semantic features, GRU focuses on capturing the sequence information of the language.
Note that LSTM generally performs comparably to GRU in sequence encoding tasks, but is more complex and require
more computational resources for training \cite{chung2014empirical}.

\subsubsection{Attentive Encoder}
An attentive encoder imports the self-attention mechanism to the recurrent neural sequence encoder, which seeks to capture the overall meaning of the input sequence unevenly from each encoded sequence item.
The self-attention is imported to GRU as below.
%\begin{align*}
%\inv
%\begin{split}~\label{eq:attention}
%&\mathbf{u}_t=\mathrm{tanh}\left (\mathbf{M}_a\mathbf{s}_t+\mathbf{b}_a\right )\\
%&a_t=\frac{\mathrm{exp}\left (\mathbf{u}^\top_t\mathbf{x}_t\right )}{\sum_{x_i\in X}\mathrm{exp}\left (\mathbf{u}^\top_i\mathbf{x}_i\right )}\\
%&\mathbf{v}_t=|X|a_t\mathbf{u}_t
%\end{split}
%\inv
%\end{align*}
\begin{align*}
\label{eq:attention}
\mathbf{u}_i=\mathrm{tanh}\left (\mathbf{M}_a\mathbf{h}^{(3)}_i+\mathbf{b}_a\right ); \qquad
a_i=\frac{\mathrm{exp}\left (\mathbf{u}^\top_i\mathbf{u}_X\right )}{\sum_{x_i\in X}\mathrm{exp}\left (\mathbf{u}^\top_i\mathbf{u}_X\right )}; \qquad
\mathbf{h}^{(4)}_i=|X|a_i\mathbf{u}_i
\end{align*}

$\mathbf{u}_i$ is the intermediary latent representation of GRU output $\mathbf{h}^{(3)}_i$, $\mathbf{u}_X=\mathrm{tanh} (\mathbf{M}_a\mathbf{h}^{(3)}_X+\mathbf{b}_a )$ is the intermediary latent representation of the averaged GRU output $\mathbf{h}^{(3)}_X$ that can be seen as a high-level representation of the entire input sequence.
By measuring the similarity of each $\mathbf{u}_i$ with $\mathbf{u}_X$, the normalized attention weight $a_i$ for $\mathbf{h}^{(3)}_i$ is produced through a softmax function.
Note that a coefficient $|X|$ (the length of the input sequence) is applied along with $a_i$ to $\mathbf{u}_i$ to obtain the weighted representation $\mathbf{h}^{(4)}_i$, so as to keep $\mathbf{h}^{(4)}_i$
from losing the original scale of $\mathbf{h}^{(3)}_i$.\par

\subsubsection{Document Embeddings}

We adopt each one of the three encoding techniques, i.e. the convolution layer with pooling, the GRU, and the attentive GRU, to form three types of document encoders respectively.
Each encoder consists of one or a stack of the corresponding layers depending on the type of the input document, and encodes the document into a embedding vector.
The document embedding is obtained by applying an affine layer to the average of the output $\mathbf{H}$ from the last pooling, GRU, or attentive GRU layer, i.e.
$E(X)=\mathbf{M}_d \left ( \frac{1}{|\mathbf{H}|}\sum^{|\mathbf{H}|}_{i=1}\mathbf{h_i}\right ) + \mathbf{b}_d$.

\subsection{Explicit Features}\label{sec:explicit_features}
In addition to the implicit semantic features provided by document encoders, we define a set of explicit features $F(A_i, A_j)=\{r_{tto}, r_{st}, r_{indeg}, r_{mt}, f_{TF}, I_{MW}, r_{outdeg}, d_{te}, r_{dt}\}$.
A portion of the explicit features are carried forward from \cite{lin2017problematizing} to provide some token-level and structural measures of an article pair $(A_i, A_j)$, which are listed as below.

\begin{itemize}%[leftmargin=1.2em]
  \item $r_{tto}$: token overlap ratio of titles, i.e. the number of overlapped words between $t_i$ and $t_j$ divided by $|t_i|$.
  \item $r_{st}$: the maximum of the token overlap ratios among the section titles of $A_i$ and those of $A_j$.
  \item $r_{indeg}$: the in-degree ratio, which is the number of incoming links in $A_i$ divided by that of $A_j$. $r_{indeg}$ measures the relative centrality of $A_i$ with regard to $A_j$.
  \item $r_{mt}$: the maximum of the token overlap ratios between any anchor title of the main-article template~\footnote{\url{https://en.wikipedia.org/wiki/Template:Main}} of $A_i$ and $t_j$, or zero if the main-article template does not apply to $A_i$.
  \item $f_{TF}$: normalized term frequency of $t_i$ in $c_j$.
  \item $d_{MW}$: Milne-Witten Index \cite{milne2008learning} of $A_i$ and $A_j$, which measures the similarity of incoming links of these two articles via the Normalized Google Distance \cite{cilibrasi2007google}.
\end{itemize}

In addition to the above features defined by \cite{lin2017problematizing}, we also include the following features.

\begin{itemize}%[leftmargin=1.2em]
  \item $r_{outdeg}$: the out-degree ratio, which measures the relative centrality of $A_i$ and $A_j$ similar to $r_{indeg}$.
  \item $d_{te}$: the average embedding distance of tokens in titles $t_i$ and $t_j$.
  \item $r_{dt}$: token overlap ratio of $c_i$ and $c_j$, which is used in \cite{lin2002single} to measure the document relatedness. In practice the calculation of $r_{dt}$ is based on the first paragraphs.
\end{itemize}

We normalize the distance and frequency-based features (i.e. $f_{TF}$ and $d_{te}$. $d_{MW}$ is already normalized by its definition.) using feature scaling \footnote{\url{https://en.wikipedia.org/wiki/Feature_scaling}}.
Note that, we do not preserve the two cross-lingual features in \cite{lin2017problematizing}. %which consider proportion of languages where the given article pairs have been identified as main and sub-articles, and the relative multilingual ``globalness'' of the main article.
This is because, in general, these two cross-lingual features are not applicable when the candidate space scales up to the range of the entire Wikipedia, since the inter-lingual links that match articles across different languages are far from complete \cite{chen2017multigraph,lehmann2015dbpedia}.

%\subsection{Annotated Word Embeddings}
%We pre-train the Skipgram \cite{mikolov2013word2vec} word embeddings on the English Wikipedia dump to support the input of the article titles and text contents to the model, as well as the calculation of the feature $d_{te}$.
%We parse all the inline hyperlinks of Wikipedia dump to the corresponding article titles, and tokenize the article titles in the plain text corpora via Trie-based maximum token matching.
%This tokenization process aims at including Wikipedia titles in the vocabulary of word embeddings, although it does not ensure all the titles to be involved as some of them occur too rarely in the corpora to meet the minimum frequency requirement of the word embedding model.
%This tokenization process is also adopted during the calculation of $d_{te}$.
%After pre-training, we fix the word embeddings to convert each document to a sequence of vectors to be fed into the document encoder.

\subsection{Training}\label{sect:train}

\begin{figure}[t]
  \centering
  \includegraphics[width=0.81\textwidth]{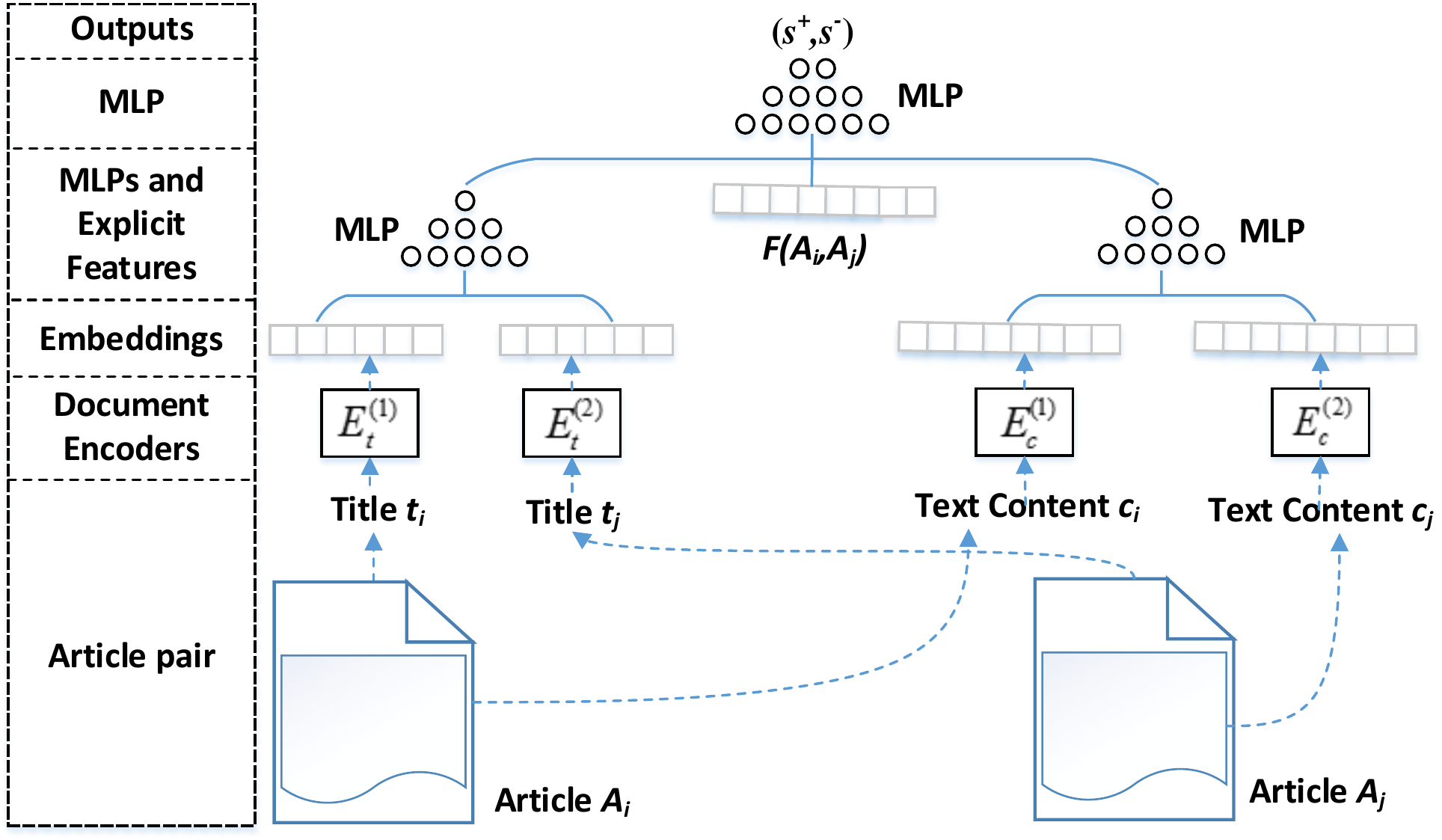}\\
  \caption{Learning architecture of the model.}\label{fig:arch}
  \vspace{-1em}
\end{figure}

\stitle {Learning objective.}
The overall learning architecture of our model is shown in Fig.~\ref{fig:arch}.
The model characterizes each given article pair $p=(A_i, A_j)\in P$ in two stages.
\begin{enumerate}%[leftmargin=1.2em]
  \item Four document encoders (of the same type) are used to encode the titles and text contents of $A_i$ and $A_j$ respectively, which are denoted as $E^{(1)}_{t}(t_i)$, $E^{(2)}_{t}(t_j)$, $E^{(1)}_{c}(c_i)$ and $E^{(2)}_{c}(c_j)$.
      Two logistic regressors realized by sigmoid multi-layer perceptrons (MLP) \cite{bengio2009learning} are applied on $E^{(1)}_{t}(t_i)\oplus E^{(2)}_{t}(t_j)$ and $E^{(1)}_{c}(c_i)\oplus E^{(2)}_{c}(c_j)$ to produce two confidence scores $s_t$ and $s_c$ for supporting $A_j$ to be the sub-article of $A_i$.
  \item The two semantic-based confidence scores are then concatenated with the explicit features ($\{s_t, s_c\} \oplus F(A_i, A_j)$), to which another linear MLP is applied to obtain the two confidence scores $\hat{s}^+_p$ and $\hat{s}^-_p$ for the boolean labels of positive prediction $l^+$ and negative prediction $l^-$ respectively. Finally, $\hat{s}^+_p$ and $\hat{s}^-_p$ are normalized by binary softmax functions $s^+_p=\frac{\mathrm{exp}\left ( \hat{s}^+_p \right )}{\mathrm{exp}(\hat{s}^+_p)+\mathrm{exp}(\hat{s}^-_p)}$ and $s^-_p=\frac{\mathrm{exp}\left ( \hat{s}^-_p \right )}{\mathrm{exp}(\hat{s}^+_p)+\mathrm{exp}(\hat{s}^-_p)}$.
\end{enumerate}

The learning objective is to minimize the following binary cross-entropy loss.

\begin{equation*}
L=-\frac{1}{|P|}\sum_{p\in P}\left ( l^+ \log s^+_p + l^-\log s^-_p \right )
\end{equation*}

%\par
\stitle{Annotated word embeddings.}
We pre-train the Skipgram \cite{mikolov2013word2vec} word embeddings on the English Wikipedia dump to support the input of the article titles and text contents to the model, as well as the calculation of the feature $d_{te}$.
We parse all the inline hyperlinks of Wikipedia dump to the corresponding article titles, and tokenize the article titles in the plain text corpora via Trie-based maximum token matching.
This tokenization process aims at including Wikipedia titles in the vocabulary of word embeddings.
Although this does not ensure all the titles to be involved, as some of them occur too rarely in the corpora to meet the minimum frequency requirement of the word embedding model.
This tokenization process is also adopted during the calculation of $d_{te}$.
After pre-training, we fix the word embeddings to convert each document to a sequence of vectors to be fed into the document encoder.

\def\hits{\mathit{Hits}\mbox{@}10}
\def\mean{\mathit{Mean}}

\section{Experiments}

In this section, we present the experimental evaluation of the proposed model.
We first create a dataset that contains a large collection of candidate article pairs for the sub-article matching problem.
Then we compare variants of the proposed model and previous approaches based on held-out estimation on the dataset.
Lastly, to show the practicability and generality of our approach, we train the model on the full dataset, and perform predictions using MapReduce on over 108 million candidate article pairs extracted from the entire English Wikipedia.

\subsection{Data Preparation}~\label{sect:data}

We have prepared a new dataset, denoted \dsnamens\footnote{\url{https://github.com/muhaochen/subarticle}}, in two stages.
We start with producing positive cases via crowdsourcing.
In detail, we first select a set of articles, where each article title concatenates two Wikipedia entity names directly or with a proposition, e.g. \emph{German Army} and \emph{Fictional Universe of Harry Potter}.
We hypothesize that such articles are more likely to be a sub-article of another Wikipedia article.
Note that this set of articles exclude the pages that belong to a meta-article category such as \emph{Lists}~\footnote{\url{https://en.wikipedia.org/wiki/Category:Lists}} and \emph{Disambiguation}~\footnote{\url{https://en.wikipedia.org/wiki/Wikipedia:Disambiguation}}, which usually do not have text contents.
Then we sample from this set of articles for annotation in the internal crowdsourcing platform of \ggnamens.
For each sampled article, we follow the criteria in Section~\ref{sect:prelim} to instruct the annotators to decide whether it is a sub-article, and to provide the URL to the corresponding main-article if so.
Each crowdsourced article has been reviewed by three annotators, and is adopted for the later population process of the dataset if total agreement is reached.
Within three months, we have obtained 17,349 positive matches of main and sub-article pairs for 5,012 main-articles, and around 4k other negative identifications of sub-articles.
\par
Based on the results of crowdsourcing, we then follow several strategies to create negative cases: (1) For each positive match $(A_i, A_j)$, we insert the inverted pair $(A_j, A_i)$ as a negative case based on the anti-symmetry of sub-article relations, therefore producing 17k negative cases; (2) For each identified main-article, if multiple positively matched sub-articles coexist, such sub-articles are paired into negative cases as they are considered as ``same-level articles''. This step contributes around 27k negative cases; (3) We substitute $A_i$ with other articles that are pointed by an inline hyperlink in $c_j$, or substitute $A_j$ with samples from the 4k negative identifications of sub-articles in stage 1. We select a portion from this large set of negative cases to ensure that each identified main-article has been paired with at least 15 negative matches of sub-articles. This step contributes the majority of negative cases. In the dataset, we also discover that around 20k negative cases are measured highly ($>0.6$) by at least one of the symbolic similarity measures $r_{tto}$, $r_{st}$ or $f_{TF}$.\par
The three strategies for negative case creation seek to populate the \dsname dataset with a large amount of negative matches of articles that represent different counterfactual cases.
The statistics of \dsname are summarized in Table~\ref{tbl:stat}, which indicate it to be much more large-scale than the dataset used by previous approaches \cite{lin2017problematizing} that contains around 3k article pairs~\footnote{Note that the text contents or URIs of participating articles that are necessary to the support of document encoding are also not provided by the 3k-article-pair dataset, thus our experiment uses \dsname only.}.
%The creation of multi-fold negative cases for each positive case is coherent in accord with the general situation of the Wikipedia where a the sub-article matching is a rare relation among article pairs.
The creation of more negative cases than positive cases is in accord with the general circumstances of the Wikipedia where the sub-article relations hold for a small portion of the article pairs.
Hence, the effectiveness of the model should be accordingly evaluated by how precisely and completely it can recognize the positive matches from all candidate pairs from the dataset.
As we have stated in Section~\ref{sect:prelim}, we encode the first paragraph of each article to represent its text contents.

\subsection{Evaluation}
We use a held-out estimation method to evaluate our approach on \dsnamens.
%~\footnote{Note that the 3k-article-pair dataset provided by \cite{lin2017problematizing} does not contain text contents of articles or URIs that help linking text contents, thus is not evaluated in our experiment.}.
Besides three proposed model variants that combine a specfic type of neural document encoders with the explicit features, we compare several statistical classification algorithms that \cite{lin2017problematizing} have trained on the explicit features.
We also compare with three neural document pair encoders without explicit features that represent the other line of related work \cite{yin2015convolutional,hu2014convolutional,sha2016reading,rocktaschel2015reasoning}.
\par
\stitle{Model configurations.}
We use AdaGrad to optimize the learning objective function and set the learning rate as 0.01, batchsize as 128.
For document encoders, we use two convolution/GRU/attentive GRU layers for titles, and two layers for the text contents.
When inputting articles to the document encoders, we remove stop words in the text contents, zero-pad short ones and truncate overlength ones to the sequence length of 100.
We also zero-pad short titles to the sequence length of 14, which is the maximum length of the original titles.
The dimensionality of document embeddings is selected among \{100, 150, 200, 300\}, for which we fix 100 for titles and 200 for text contents.
For convolutional encoders, we select the kernel size and the pool size from 2 to 4, with the kernel size of 3 and 2-max-pooling adopted.
For pre-trained word embeddings, we use context size of 20, minimum word frequency of 7 and negative sampling size of 5 to obtain 120-dimensional embeddings from the tokenized Wikipedia corpora mentioned in Section~\ref{sect:train}.
Following the convention \cite{yilmaz2011multiple,bengio2009learning}, we use one hidden layer in MLPs where the hidden-layer size is the average of the input and output layer sizes.\par
\stitle{Evaluation protocal.}
Following \cite{lin2017problematizing}, we adopt 10-fold cross-validation in the evaluation process.
At each fold, all models are trained till converge.
We aggregate \emph{precision}, \emph{recall} and \emph{F1-score} on the positive cases at each fold of testing,
since the objective of the task is to effectively identify the relatively rare article relation among a large number of article pairs.
All three metrics are preferred to be higher to indicate better performance.\par
%
%\begin{wraptable}{r}{0.33\textwidth}
%\setlength\tabcolsep{2pt}
%\caption{Statistics of the dataset.}
%\label{tbl:stat}
%\begin{tabular}{c|c}
%\bhline
%\#Article pairs&195,960\\
%\#Positive cases&17,349\\
%\#Negative cases&178,611\\
%\#Main-articles&5,012\\
%\#Distinct articles&32,487\\
%\bhline
%\end{tabular}
%\end{wraptable}

\begin{table}[t]
\setlength\tabcolsep{2pt}
\centering
\caption{Statistics of the dataset.}
\label{tbl:stat}
\small
\begin{tabular}{c|c|c|c|c}
\bhline
\#Article pairs&\#Positive cases&\#Negative cases&\#Main-articles&\#Distinct articles\\
\hline
195,960&17,349&178,611&5,012&32,487\\
\bhline
\end{tabular}
\end{table}

\begin{table}[t]
\setlength\tabcolsep{2pt}
%\scriptsize
\centering
\caption{Cross-validation results on \dsnamens. We report precision, recall and F1-scores on three groups of models: (1) statistical classification algorithms based on explicit features, including logistic regression, Naive Bayes Classifier (NBC), Linear SVM, Adaboost (SAMME.R algorithm \cite{hastie2009multi}), Decision Tree (DT), Random Forest (RF) and k-nearest-neighbor classifier (kNN); (2) three types of document pair encoders without explicit features; (3) the proposed model in this paper that combines explicit features with convolutional document pair encoders (CNN+$F$), GRU encoders (GRU+$F$) or attentive encoders (AGRU+$F$).}
\label{tbl:cv}
\small
\begin{tabular}{c|ccccccc}
\bhline
\multirow{2}{*}{Model}&\multicolumn{7}{c}{Explicit Features}\\
\cline{2-8}
&Logistic&NBC&Adaboost&LinearSVM&DT&RF&kNN\\
\hline
Precision (\%)&82.64&61.78&87.14&82.79&87.17&89.22&65.80\\
%\hline
Recall (\%)&88.41&87.75&85.40&89.56&84.53&84.49&78.66\\
%\hline
F1-score&0.854&0.680&0.863&0.860&0.858&0.868&0.717\\
\hline
\multirow{2}{*}{Model}&\multicolumn{3}{c|}{Semantic Features}&\multirow{2}{*}{Model}&\multicolumn{3}{|c}{Explicit+Semantic}\\
\cline{2-4}
\cline{6-8}
&CNN&GRU&\multicolumn{1}{c|}{AGRU}&&\multicolumn{1}{|c}{CNN+$F$}&GRU+$F$&AGRU+$F$\\
\hline
Precision (\%)&95.83&95.76&\multicolumn{1}{c|}{93.98}&Precision (\%)&\multicolumn{1}{|c}{\textbf{99.13}}&98.60&97.58\\
%\hline
Recall (\%)&90.46&87.24&\multicolumn{1}{c|}{86.47}&Recall (\%)&\multicolumn{1}{|c}{\textbf{98.06}}&88.47&86.80\\
%\hline
F1-score&0.931&0.913&\multicolumn{1}{c|}{0.901}&F1-score&\multicolumn{1}{|c}{\textbf{0.986}}&0.926&0.919\\
\bhline
\end{tabular}
%\vspace{-1em}
\end{table}

%\begin{wraptable}{r}{0.39\textwidth}
\begin{table*}[t]
\begin{minipage}[h]{1.\textwidth}
\setlength\tabcolsep{1pt}
\centering
\captionsetup{justification=centering}
\caption{Ablation on feature categories for CNN+$F$.}
%\vspace{-1em}
\label{tbl:ab}
\small
\begin{tabular}{l|ccc}
\bhline
Features&Precision&Recall&F1-score\\
\hline
All features&99.13&98.06&0.986\\
%\hline
No titles&98.03&85.96&0.916\\
%\hline
No text contents&98.55&95.78&0.972\\
%\hline
No explicit&95.83&90.46&0.931\\
Explicit only&82.64&88.41&0.854\\
\bhline
\end{tabular}
\end{minipage}
\hfill
\begin{minipage}[h]{1.\textwidth}
  \centering
  \includegraphics[width=0.7\textwidth]{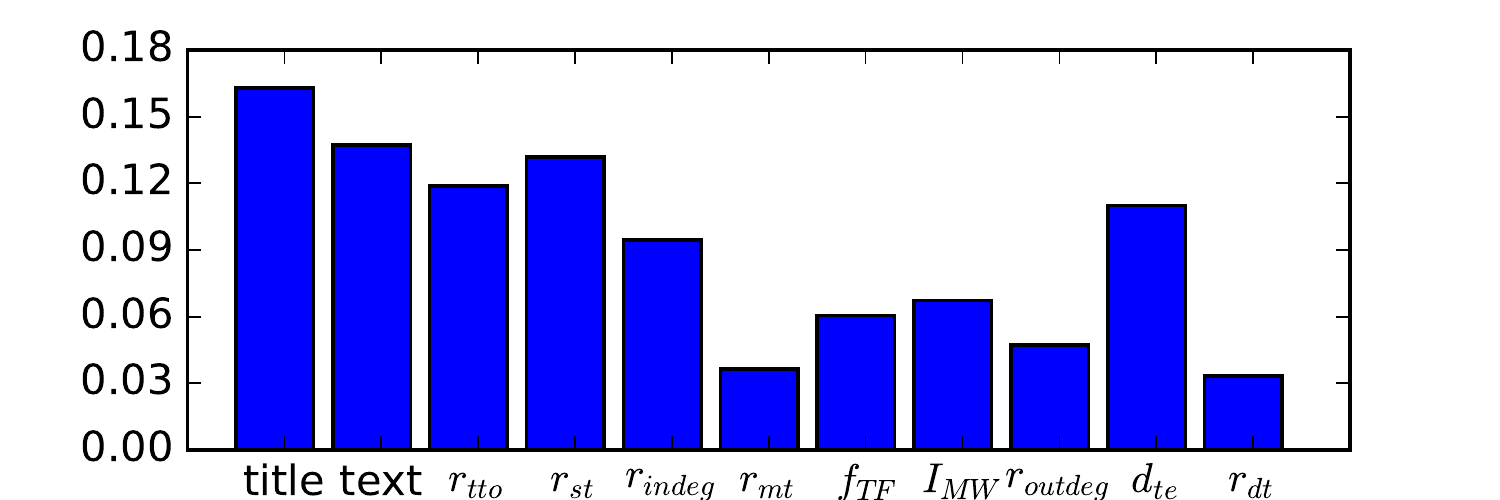}\\
  \captionsetup{justification=justified}
  \captionof{figure}{Relative importance (RI) of features analyzed by Garson's algorithm. RI of each feature is aggregated from all folds of cross-validation.}\label{fig:ri}
\end{minipage}
\hfill
\begin{minipage}[h]{1.\textwidth}
  \centering
  \includegraphics[width=0.7\textwidth]{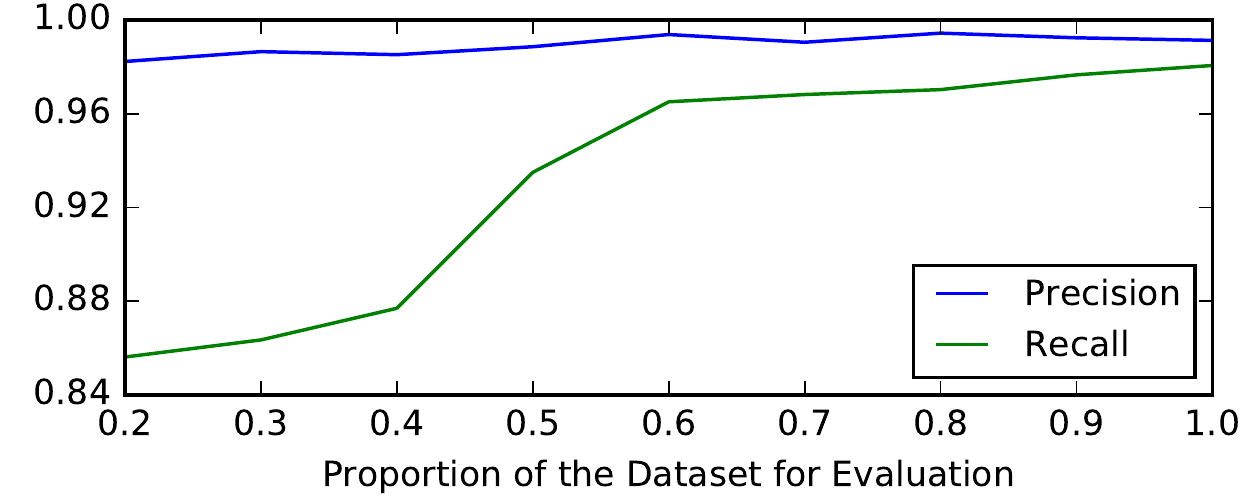}\\
  \captionsetup{justification=centering}
  \captionof{figure}{Sensitivity of CNN+$F$ on the proportion of dataset for evaluation.}\label{fig:size}
  \vspace{-2em}
\end{minipage}
\end{table*} 
\stitle{Results.} Results are reported in Table~\ref{tbl:cv}.
The explicit features alone are helpful to the task, on which the result by the best baseline (Random Forest) is satisfactory.
However, the neural encoders for document pairs, even without the explicit features, outperform Random Forest by 4.76\% on precision, 1.98\% on recall and 0.033 on F1-score.
This indicates that the implicit semantic features are critical for characterizing the matching of main and sub-articles.
Among the three types of document encoders, the convolutional encoder is more competent than the rest two sequence encoders, which outperforms Random Forest by 6.54\% of precision, 8.97\% of recall and 0.063 of F1-score.
This indicates that the convolutional and pooling layers that effectively capture the local semantic features are key to the identification of sub-article relations, while such relations appear to be relatively less determined by the sequence information and overall document meanings that are leveraged by the GRU and attention encoders.
The results by the proposed model which combines document pair encoders and explicit features are very promising.
Among these, the model variant with convolutional encoders (CNN+$F$) obtained close to perfect precision and recall.\par
Meanwhile, we perform ablation on different categories of features and each specific feature, so as to understand their significance to the task.
Table~\ref{tbl:ab} presents the ablation of feature categories for the CNN-based model.
We have already shown that completely removing the implicit semantic features would noticeably impair the precision.
Removing the explicit features moderately hinders both precision and recall.
As for the two categories of semantic features, we find that removing  either of them would noticeably impair the model performance in terms of recall, though the removal of title embeddings has much more impact than that of text content embeddings.
Next, we perform Garson's algorithm~\cite{olden2002illuminating,feraud2002methodology} on the weights of the last linear MLP of CNN+$F$ to analyze the relative importance (RI) of each specific feature, which are reported as Fig.~\ref{fig:ri}.
It is noteworthy that, besides the text features, the explicit features $r_{tto}$, $r_{st}$ and $d_{te}$ that are related to article or section titles also show high RI.
This is also close to the practice of human cognition, as we humans are more likely to be able to determine the semantic relation of a given pair of articles based on the semantic relation of the titles and section titles than based on other aspects of the explicit features.
\par
Furthermore, to examine how much our approach may rely on the large dataset to obtain a generic solution,
we conduct a sensitivity analysis of CNN+$F$ on the proportion of the dataset used for cross-validation, which is reported in Fig.~\ref{fig:size}.
We discover that, training the model on smaller portions of the dataset would decrease the recall of predictions by the model, though the impact on the precision is very limited.
However, even using 20\% of the data, CNN+$F$ still obtains better precision and recall than the best baseline Random Forest that is trained solely on explicit features in the setting of full dataset.
\par
To summarize, the held-out estimation on the \dsname dataset shows that the proposed model is very promising in addressing the sub-article matching task.
Considering the large size and heterogeneity of the dataset, we believe the best model variant CNN+$F$ is close to a well-generalized solution to this problem.
\par
\subsection{Mining Sub-articles from the Entire English Wikipedia}
\begin{table}[t]
\setlength\tabcolsep{2pt}
\centering
\caption{Examples of recognized main and sub-article matches. The italicize sub-article titles are without overlapping tokens with the main article titles.}
\label{tbl:case}
%\vspace{-1em}
{
\scriptsize
\begin{tabular}{l|l}
\bhline
Main Article&Sub-articles\\
\bhline
Outline of government&\emph{Bicameralism}, \emph{Capitalism}, \emph{Dictatorship}, \emph{Confederation}, \emph{Oligarchy}, \emph{Sovereign state}\\
\hline
\multirow{2}{*}{Computer}&Computer for operations with functions, Glossary of computer hardware terms, Computer\\
&user, \emph{Timeline of numerical analysis after 1945}, Stored-program computer, Ternary computer\\
\hline
Hebrew alphabet&Romanization of Hebrew\\
\hline
Recycling by material&Drug recycling, \emph{Copper}, \emph{Aluminium}, Drug recycling\\
\hline
\multirow{2}{*}{Chinese Americans}&History of Chinese Americans in Dallas-Fort Worth, History of Chinese Americans in San\\
&Francisco, Anti-Chinese Violence in Washington\\
\hline
Genetics&Modification (Genetics), Theoretical and Applied Genetics, Encyclopedia of Genetics\\
\hline
\multirow{3}{*}{San Marino}&Economy of San Marino, San Marino national football team, Democratic Convention (San\\
&Marino), Banca di San Marino, Healthcare in San Marino, Flag of San Marino, Geography of\\
&San Marino\\
\hline
\multirow{2}{*}{Service Rifle}&United States Marine Corps Squad Advanced Marksman Rifle, United States Army Squad\\
&Designated Marksman Rifle\\
\hline
\multirow{3}{*}{Transgender rights}&LGBT rights in Panama, LGBT rights in the United Arab Emirates, Transgender rights in\\
&Argentina, History of transgender people in the United States, Transgender disenfranchisement\\
&in the United States\\
\hline
Spectrin&Spectrin Repeat\\
\hline
Geography&Political Geography, Urban geography, Visual geography, \emph{Colorado Model Content Standards}\\
\hline
Nuclear Explosion&Outline of Nuclear Technology, International Day Against Nuclear Tests\\
\hline
Gay&\emph{LGBT Rights by Country or Territory}, Philadelphia Gay News, Troll (gay slang), Gay literature\\
\hline
FIBA Hall of Fame&\emph{\u{S}ar\={u}nas Mar\u{c}iulionis}\\
\hline
Arve Isdal&\emph{March of the Norse}, \emph{Between Two Worlds}\\
\hline
Independent politician&\emph{Balasore (Odisha Vidhan Sabha Constituency)}\\
\hline
\multirow{4}{*}{Mathematics}&Hierarchy (mathematics), \emph{Principle part}, Mathematics and Mechanics of Complex Systems,\\
&\emph{Nemytskii operator}, \emph{Spinors in three dimensions}, \emph{Continuous functional calculus}, Quadrature,\\
&Table of mathematical symbols by introduction date, \emph{Hasse invariant of an algebra}, Concrete\\
&Mathematics\\
\hline
Homosexuality&\emph{LGBT rights in Luxembourg}, List of Christian denominational positions on homosexuality\\
\hline
Bishop&\emph{Roman Catholic Diocese of Purnea}, \emph{Roman Catholic Diocese of Luoyang}\\
\hline
Lie algebra&Radical of a Lie algebra, Restricted Lie algebra, \emph{Adjoint representation}, Lie Group\\
\bhline
\end{tabular}
}
%\vspace{-1em}
\end{table}

For the next step, we move on to putting the proposed model into production by serving it to identify the main and sub-article matching on the entire body of the English Wikipedia.
The English Wikipedia contains over 5 million articles, which lead to over 24 trillion ordered article pairs.
Hence, instead of serving our model on that astronomical candidate space, we simplify the task by predicting only for each article pair that forms an inline hyperlink across Wikipedia pages, except for those that appear already in the \emph{main-article} templates.
This reduces our candidate space to about 108 million article pairs.
\par
We train the best model variant CNN+$F$ from the previous experiment for serving.
We carry forward the model configurations from the previous experiment.
The model is trained on the entire \dsname dataset till converge.
The extraction of the candidate article pairs as well as the serving of the model is conducted via MapReduce on 3,000 machines, which lasts around 9 hours in total.
We select the 200,000 positive predictions with highest confidence scores $s^+_p$, based on which human evaluation on three turns of 1,000 sampled results estimates a 85.7\% of $P@200k$ (precision at top 200,000 predictions).
Examples of identified main and sub-article matches are listed in Table~\ref{tbl:case}.
Based on the selected positive predictions, the number of sub-articles per main-article is estimated as 4.9, which is lower than 7.5 that is estimated on the 1,000 most viewed articles by \cite{lin2017problematizing}.
There are also around 8\% of sub-articles that are paired with more than one main-articles.
Based on the promising results from the large-scale model serving, our team is currently working on populating the identified sub-article relations into the backend knowledge base for our search engine.

\section{Conclusion and Future Work}

In this paper, we have proposed a neural article pair model to address the sub-article matching problem in \mbox{Wikipedia}.
The proposed model utilizes neural document encoders for titles and text contents to capture the latent semantic features from Wikipedia articles,
for which three types of document encoders have been considered, including the convolutional, GRU and attentive encoders.
A set of explicit features are incorporated into the learning framework that comprehensively measured the symbolic and structural similarity of article pairs.
We have created a large article pair dataset \dsname from English Wikipedia which seeks to generalize the problem with various patterns of training cases.
The experimental evaluation on \dsname based on cross-validation shows that the document encoders alone are able to outperform the previous models using only explicit features, while the combined model based on both implicit and explicit features is able to achieve near-perfect precision and recall.
Large-scale serving conducted on the entire English Wikipedia is able to produce a large amount of new main and sub-article matches with promising quality.
For future work, it is natural to apply the proposed model to other language-versions of Wikipedia for production.
It is also meaningful to develop an approach to differentiate the sub-articles that describe refined entities and those that describe abstract sub-concepts.
Meanwhile, we are interested in extending our approach to populate the incomplete cross-lingual alignment of Wikipedia articles using bilingual word embeddings such as Bilbowa \cite{gouws2015bilbowa}, and a different set of explicit features.

%\clearpage
\bibliographystyle{splncs04}
\begingroup
%\setstretch{0.85}
\bibliography{ref}
\endgroup

\end{document}